\documentclass[times,onecolumn]{article}

\usepackage[top=2cm,left=2cm,right=2.5cm,bottom=2 cm,includeheadfoot]{geometry}

\usepackage{authblk}
\usepackage{framed,multirow}
\usepackage{amssymb}
\usepackage{latexsym}
\usepackage{url}
\usepackage{xcolor}
\usepackage{graphicx} 
\usepackage{caption}
\usepackage{epstopdf}
\usepackage{amsmath} 
\usepackage{float}
\usepackage{threeparttable} 
\usepackage{comment}

\definecolor{newcolor}{rgb}{.8,.349,.1}

\begin{document}

\title{\bf Classification of Lung Pathologies in Neonates using Dual Tree Complex Wavelet Transform}

\author{Sagarjit Aujla $^{1}$, Adel Mohamed $^{2}$, Ryan Tan $^{1}$, Randy Tan $^{1}$, Lei Gao  $^{1}$, Naimul Khan $^{1}$, and Karthikeyan Umapathy}

\affil[1]{Toronto Metropolitan University, 350 Victoria St, Toronto M5B 2K3, Canada}
\affil[2]{Mount Sinai Hospital, 600 University Ave, Toronto M5G 1X5, Canada}

\maketitle

\begin{abstract}

Annually 8500 neonatal deaths are reported in the US due to respiratory failure, this is a third of all neonatal deaths in the US. Recently, Lung Ultrasound (LUS), due to its ionizing-radiation free nature, portability, and being cheaper is gaining wide acceptability as a diagnostic tool for different lung conditions. However, lack of highly trained medical professionals has limited its use especially in remote areas. To address this, an automated screening system that captures characteristics of the LUS patterns can be of significant assistance to clinicians who are not experts in lung ultrasound (LUS) images. In this paper, we propose a feature extraction method designed to quantify the spatially-localized line patterns and texture patterns found in LUS images. Using the  dual-tree complex wavelet transform (DTCWT) and four types of common image features we propose a method to classify the LUS images into 6 common neonatal lung conditions. These conditions are normal lung, pneumothorax (PTX), transient tachypnea of the newborn (TTN), respiratory distress syndrome (RDS), chronic lung disease (CLD) and consolidation (CON) that could be pneumonia or atelectasis. The proposed method using DTCWT decomposition extracted global statistical, grey-level co-occurrence matrix (GLCM), grey-level run length matrix (GLRLM) and linear binary pattern (LBP) features to be fed to a linear discriminative analysis (LDA) based classifier. Using 15 best DTCWT features along with 3 clinical features the proposed approach achieved a per-image classification accuracy of {\bf 92.78\%} with a balanced dataset containing 720 images from 24 patients and {\bf 74.39\%} with the larger (class unbalanced) dataset containing 1550 images from 42 patients. Likewise, the proposed method achieved a maximum per-subject classification accuracy of {\bf 81.53\%} with 43 DTCWT features and 3 clinical features using the balanced dataset and {\bf 64.97\%} with 13 DTCWT features and 3 clinical features using the unbalanced larger dataset.

\end{abstract}

\section{INTRODUCTION}

\begin{figure*}
	\centering
	\captionsetup{justification=centering}
	\includegraphics[scale=0.60]{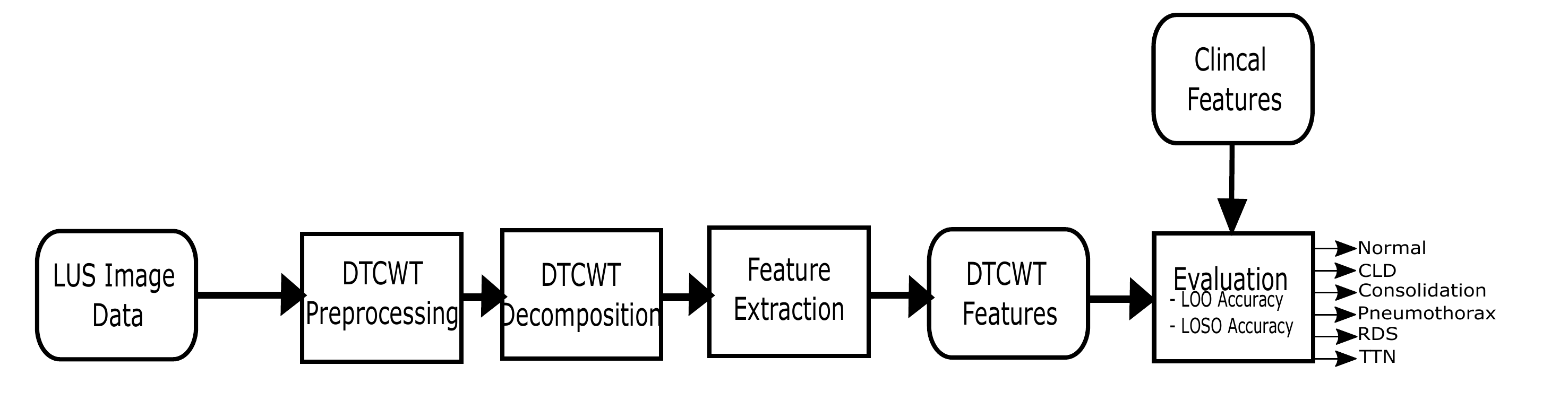}
	\vspace{-0.0in}
	\caption{Block diagram of the proposed method}
	\vspace{-0.0in}
	\label{figure1}
\end{figure*}

The use of lung ultrasound (LUS) to diagnose and monitor lung pathologies in neonates has increased in many urban hospitals. This can be attributed to the advantages that it has over conventional imaging modalities, X-rays and CT scans. Compared to X-rays and CT scans, LUS is cheaper, more accessible and ionizing-radiation free. Furthermore, LUS is comparable to bedside chest X-ray and chest CT in diagnosing lung diseases \cite{Comparison},\cite{c6}. Specialist medical professionals and clinicians are required to carry out the LUS scans on neonates and read the LUS scans to determine the lung condition of the neonates. There is lack of these highly trained medical professionals and clinicians in hospitals in developing countries and remote hospitals. In these environments, an automated classification tool could be used to assist medical professionals and physicians who use LUS as a diagnostic tool.

Attempts to automate the diagnosis of specific clinical disorders from LUS have already been made. In \cite{diaz2021deep} and \cite{la2021deep}, the researchers detect Covid-19 pneumonia using deep learning approaches. In Tsai et al. \cite{c7}, the authors performed binary classification of pleural effusion using a convolutional neural network. In the following works, features related to the common clinical LUS morphologies were extracted and used. In \cite{Key_Features}, pleural lines, A-lines and B-lines were extracted and used for classification. In \cite{c8}, researchers detected the pleural line in the image and selected a region of interest (ROI) below the pleural line to grade the severity of Covid pneumonia. In \cite{carrer2020automatic}, researchers detected the pleural line and extracted features related to the pleural line to evaluate the healthiness of the patient. In Correa et al. \cite{c9}, the researchers extracted thin rectangular regions of the image and used these as inputs to a 3-layer feed-forward neural network to classify pediatric pneumonia. The majority of the existing works in general take into account the primary morphologies for each individual pathological condition they are intended to detect. To distinguish between various lung diseases, it is necessary to consider all of the major LUS morphologies connected to these disorders. In Bassiouny et al \cite{c10}, 7 key LUS morphologies that are key markers for specific lung conditions  were detected using a Faster Region-Based Convolutional Neural Network (FRCNN) object detection model.

In our previous initial work \cite{RQA_Paper}, we proposed a feature extraction method that was designed to quantify the strong recurrence characteristics of the image morphologies used by doctors in diagnosing LUS images. We extracted scanlines from the images, these scanlines were shown to capture the commonly used LUS morphologies for diagnosis and used recurrence analysis to capture nonlinear features based on recurrence patterns captured in the scanlines. We had also included 3 clinical features viz. gestational age (GA), corrected gestational age at the time of the scan (CGATS), and day(s) of life (DOL). These features are required to separate CLD and RDS as advised by our clinical collaborators.

\begin{figure*}[!h]
	\centering
	\includegraphics[scale=0.45]{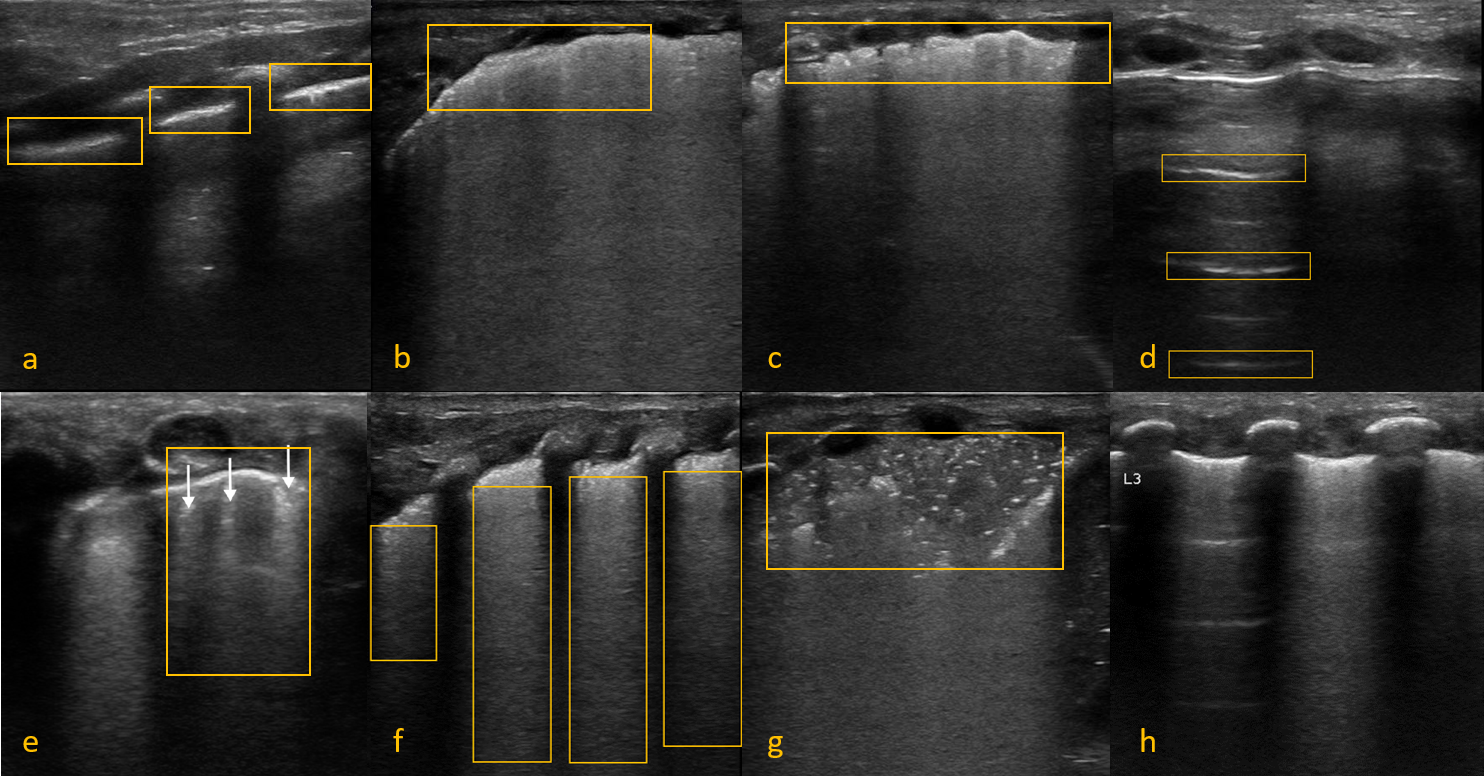}
	\caption{a: A sample normal Pleura from a subject with Normal Lung, b: A sample thick Pleura ($>$ 2mm) from a subject with CON, c: A sample thick and irregular Pleura from a subject with CLD, d: Sample A-Lines from a subject with PTX, e: Sample separate B-Lines from a subject with TTN, f: Sample coalescent B-lines from a subject with RDS, g: Sample CON illustration from a subject with Consolidation, and h: Sample Double Lung Point from a subject with TTN}
	\label{figure2}
\end{figure*}

There exists some LUS works that use approaches similar to our proposed method in this paper. In \cite{LUS_2D_Wavelet} a similar wavelet transform was used to decompose areas in the LUS images. Contreras-Ojeda et al. selected 2 regions in the LUS image corresponding to chest tissue and lung tissue and decomposed these regions into the detail coefficients using the wavelet transform. Statistical features were extracted from these regions such as energy, mean, median, standard deviation, covariance, kurtosis, root means square (RMS), and the peak-magnitude-RMS. Finally, they used K-nearest neighbors (KNN) and tested their accuracy using  10-fold cross-validation to classify the regions as muscular tissue or lung tissue. In \cite{LUS_radiomics},  Cao et al. manually selected a region of interest (ROI) containing B-lines or white lung. Then, they extracted radiomic features such as first-order statistical features using wavelet filtering, Grey Level Co-Occurrence Matrix (GLCM) features, Gray Level Run Length Matrix (GLRLM) features, Gray Level Size Zone Matrix (GLSZM) features, Neighboring Gray Tone Difference Matrix (NGTDM) features, Gray Level Dependence Matrix (GLDM) features and Shape features. Finally, they  used a support vector machine (SVM) for classifying the manually selected regions as B-line or white lung. Fundamentally, they are selecting a ROI in the image and classifying it as chest or lung in \cite{LUS_2D_Wavelet} and in \cite{LUS_radiomics} they are classifying the ROI as 2 morphologies, B-lines or white lung (which is a large area of coalescent B-lines). 

In our work, we perform a 6-class pathological condition classification by dividing our image into two halves, top and bottom using similar features to the existing works on the decomposed wavelet images. We also used 3 clinical features as explained earlier to separate between pathological conditions with similar morphologies. The 6 conditions are separated by quantifying characteristics of the image morphologies associated with the conditions using a time-frequency transformation approach. This approach will be used to isolate the spatially-localized line patterns and texture patterns found in LUS. In the paper by G. Chen \cite{medical_wavelet}, the 1-D dual-tree complex wavelet transform (DTCWT) was used to extract features for automatic seizure detection from EEG signals. In the paper by D.B. Aydogan et al. \cite{image_medical_wavelet}, the 2-D DTCWT is used for the detection of bone fractures, classification and segmentation of brain tumors from MRI images. In \cite{image_medical_wavelet2}, DTCWT is used to extract features for breast tumor classification from MRI images. In \cite{Ultrasound_DTCWT}, researchers used the DTCWT to classify ultrasound images of thyroids into 6 categories. These works indicate that the DTCWT’s approximate high directional selectivity, scale-invariance, good space-frequency localization and shift invariance properties makes it an excellent time-frequency transformation algorithm for medical image analysis tasks such as the proposed LUS image classification. In LUS, the main lung morphologies are spatially isolated high intensity patterns or texture patterns. In our work, the DTCWT is used to decompose the image into subimages from which we extract a combination of textural and morphological features using global statistical pixel intensity distribution, grey-level co-occurrence matrix, grey-level run length matrix and linear binary pattern features. 

The rest of the paper is organized as follows: Section II describes the LUS morphologies in neonates, Section III defines the datasets, details of methods are presented in Section IV, results and discussion is presented in Section V, and our conclusions with future works are presented in Section VI. A block diagram outlining the proposed method is illustrated in Fig. \ref{figure1}.

\section{LUS MORPHOLOGIES IN NEONATES}

This section contains descriptions of the most important clinical markers used by clinicians in diagnosing different lung conditions as presented in our previous work \cite{RQA_Paper}. These clinical markers are Pleural lines, A-lines, Separate B-lines, Coalescent B-lines and Consolidations. Capturing the characteristics of these clinical markers is essential to successfully automate the diagnosis of LUS. 

\subsection{{\bf Pleural Line} [Fig. \ref{figure2}(a-c)]:} Identifying and characterizing pleural line artifact is the most important LUS markers for clinicians to diagnose different lung pathologies. The pleura is made up of the visceral pleura lining, that is a thin membrane which is affixed to the surface of the lungs and the parietal pleura lining, that is the membrane affixed to the chest wall. Between the two pleural layers is a thin layer of fluid that allows the layers to slide freely during respiration. In LUS, at the interface separating the pleura and the lung tissue the ultrasound waves are reflected back producing a bright (echogenic) horizontal line. This is called pleural line artifact.

\subsection{{\bf A- Lines} [Fig. \ref{figure2}(d)]:} A-lines are horizontal artifacts generated by the repeated reflection of the ultrasound beam between the pleural line and the probe surface. It only occurs when the lung is well aerated or in the case of PTX. They are typically at equal distance similar in appearance to the pleural line and generally are a sign of lung aeration.  

\subsection{{\bf B- Lines} [Fig. \ref{figure2}(e-f)]:} B-lines are a sign that there is fluid or a lack of air in the interstitial  and/or alveolar spaces of the lungs. The concentration of B-lines is correlated with the amount of fluid in the lungs. B-lines appear as vertical reverberation line artifacts. It is only a true B-line if it starts at the pleural line and extends to the bottom of the screen. A couple of B-lines can occur under normal circumstances especially in neonates as small amounts of fluid may still be present in the lungs after birth. However, $\geq$ 3 B-Lines in a LUS frame is an indication of interstitial lung disease. As they increase in number, the B-lines appear to diffuse together and are called coalescent B-lines which is an indication of alveolar-interstitial disease. 

\subsection{{\bf Consolidations} [Fig. \ref{figure2}(g)]:} Due to severe lack of aeration in the lungs, consolidations appear as hypoechoic areas with hyperechoic short lines (air-bronchogram) with irregular or absent pleural lines. 

\subsection{{\bf Double Lung Point} [Fig. \ref{figure2}(h)]:} Double Lung Point is specific to TTN and has a sensitivity of 100\% \cite{Double_Lung_Point}. The difference in aeration between the upper lung regions (A-lines) and lower lung region (B-lines) causes Double Lung Point. Absence of Double Lung Point doesn't rule out TTN.\\

\begin{figure}[!h]
	\centering
	\includegraphics[scale=0.35]{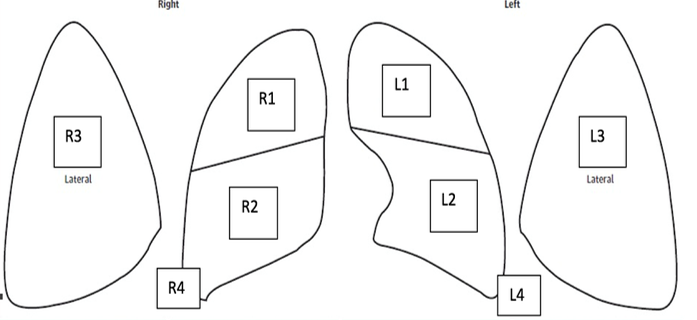}
	\caption{The 6 standard lung regions scanned during lung ultrasound (L1, L2, L3, R1, R2, R3). This is a standard clinical method.}
	\vspace{-0.2in}
	\label{figure3}
\end{figure}

In Fig. \ref{figure3} we have illustrated the six anatomical regions (3 each on the left and right side of the lungs) of a clinically standard LUS scan and in Table \ref{table1}, we have provided a brief summary of expected morphological characteristics in LUS images to assign one of the 6 lung conditions. 

\begin{table}[!h]
	\caption{Lung Conditions and the Associated Morphologies}
	\label{table1}
	\centering
	\begin{tabular}{|p{1.5cm}|p{6.5cm}|}
		\hline
		\textbf{Name} & \textbf{Morphologies}                                                                                                                                                                                                                     \\ \hline
		Normal             & Presence of A-lines in all lung regions, normal Pleural line and lung sliding.                                                                                             \\ \hline
		
		PTX                & No lung sliding in areas of PTX, but will look like a normal lung. However, no US waves pass the air between the pleural layers.\\ \hline
		
		TTN                & Normal pleural line, with indications of interstitial (separate B-lines) or alveolar-interstitial syndrome (coalescent B-lines) in the lower lung regions. Will contain A-lines in the upper lung regions. \\ \hline
		
		RDS                & Irregular and thickened pleural line and with consolidations in some areas of the lung. Will have coalescent B-line to $\geq $3 separate B-lines in all regions of the lungs. \\ \hline
		
		CON      & Will have areas of consolidation. In severe cases, the consolidated area will look like the liver tissue which is known as "hepatization".\\ \hline
		CLD                & Irregular and thickened pleural line. In addition, there may be presence of different severities of B-lines from spared areas to separate coalescent B-lines or even some spared areas. \\ \hline 
	\end{tabular}
	
\end{table}

\section{DATASET}

Our research collaborators from a Canadian tertiary neonatal intensive care unit in Mount Sinai Hospital (MSH) acquired all of the LUS scan videos. Research Ethics Board approvals and data sharing agreement were obtained from both Toronto Metropolitan University and MSH. A breakdown of the number of patients, videos and images for each condition are shown in Table \ref{dataset}. Two datasets were used for this work, a whole dataset (class unbalanced) and a balanced dataset.  The whole dataset was created using 5 frames from each of the videos acquired by our clinical collaborators. The whole dataset contained 6 or more videos for most of the patients in the dataset. For most of the patients, 1 or more videos of each of the commonly imaged areas of the lungs (R1, R2, R3, L1, L2 and L3 as shown in Figure \ref{figure3}) were included in the dataset. R4, L4 regions are imaged for specific conditions, but not for any of the conditions in the dataset. For patients with PTX, as it happens in a specific area of the lung pleura and not in both lungs, only lung areas with pneumothorax were included so these patients usually had less than 6 videos. For the whole dataset, 5 frames were taken at equal intervals in the video unless a particular frame was not clear. This was done to avoid the large similarity between neighboring frames that would bias the model. Looking at the distribution of the whole dataset in Table \ref{dataset} it is clear that the dataset is biased towards TTN, which comprises more than a third of the dataset. The second dataset, the balanced dataset, consisted of only 4 patients per condition. Six videos, one for each of the lung regions, were taken for each patient and 5 frames taken at equal intervals were used from each video. For 2 of the PTX patients, only 3 and 4 videos were available so 10 and 7 or 8 frames were taken from those videos to have 30 images per patient. The patients and videos for the balanced dataset were determined by our clinical collaborators. The balanced dataset was created to test the performance of the DTCWT features without any effects of bias stemming from a class unbalanced dataset. 

\begin{table}[]
	\centering
	\caption{Whole Dataset Overview}
	\begin{tabular}{|c|c|c|c|}
		\hline
		& Patients & Videos & Images \\ \hline
		Normal & 6        & 37     & 185    \\ \hline
		CLD    & 6        & 36     & 180    \\ \hline
		CON    & 7       & 61    & 305    \\ \hline
		PTX    & 5        & 21     & 105    \\ \hline
		RDS    & 7        & 49     & 245    \\ \hline
		TTN    & 11        & 106     & 530    \\ \hline
		\textbf{Total}  & \textbf{42}       & \textbf{310}    & \textbf{1550}   \\ \hline
	\end{tabular}
	\label{dataset}
\end{table}

\section{METHODS}

In this section, the methods and process for obtaining the DTCWT features and performing pattern classifications are presented.

\subsection{Preprocessing}

The preprocessing of the LUS images consisted of two main operations, artifact removal and normalization. The LUS videos have region information and other artifacts overlaid on the videos, which was close to the max intensity in the image. These artifacts remain in the same x-y coordinates throughout the video so these artifacts were removed semi-automatically by selecting a region of interest (ROI) over the artifact, this ROI was used for all the images in a video. The artifacts were removed by selecting all the pixels in the ROI that had an intensity greater than 50\% of the maximum intensity of the ROI. The 8-pixel connected neighbors were also selected. All the selected pixels were replaced with median intensity for the ROI. For normalization, the images were all resized to [520 420], then 10 pixels were removed from each side of the image resulting in all the images being resized to [500 400]. These steps were taken to remove the high-intensity artifacts that can easily be picked up by some of the subbands and normalize the images for DTCWT decomposition. The images were cropped to remove parts of the image that were not important and to remove some artifacts that appear at the edges of the video. 

\subsection{Time-Frequency/ Time-Scale Transformation}

The LUS morphologies can be described as a combination of information that is localized spatially as well as texture patterns and small oscillations. These patterns can be characterized as belonging to different spatial frequencies and relative locations, so a 2D time-frequency/time-scale transformations may be able to capture these patterns making it easier to extract features that can quantify these patterns. 

\subsubsection{DWT}

The 2-D discrete wavelet transform (DWT) can be used to analyze an image and isolate these different frequencies using different scales. The 2-D discrete wavelet transform uses basis functions to decompose various scaled versions of the input image \cite{Wavelet_Textbook} \cite{DWT_Filters}. The 2-D DWT is implemented by using 1D wavelet and scaling functions \cite{Wavelet_Textbook}. The 1D DWT of a signal can be defined as:
\begin{equation}
	x[n]=\sum_{k}{a_{J,k}2^{-J/2}\phi(2^{-J}n-k)}+\sum_{j,k}{d_{j,k}2^{-j/2}\psi(2^{-j}n-k)}
\end{equation}
Where $a_{J,k}$ is the approximation coefficients and $d_{j,k}$ is the detail coefficients at octave decomposition levels $j$. $\psi(n)$ is the orthonormal wavelet function, $\phi(n)$ is the scaling function and $k$ is the translation parameter. 

\begin{figure*}[]
	\centering
	\includegraphics[scale=0.6]{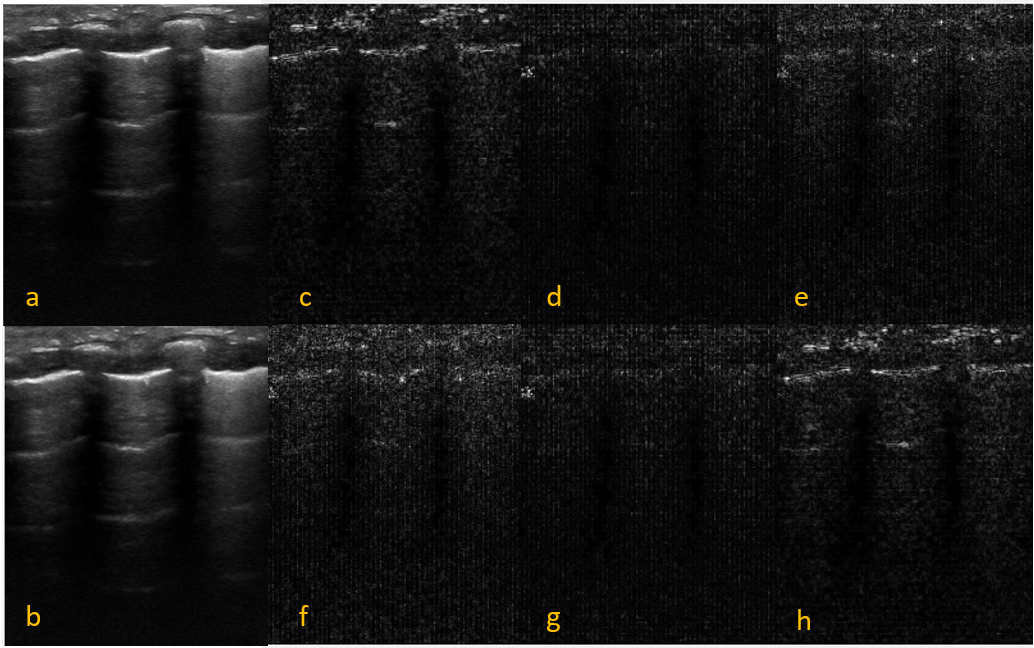}
	\caption{One level of DTCWT decomposition. a: Image of Patient with TTN, b: Low-Passed Band: smoothed image c: Subimage at +15$^{\circ}$, d: Subimage at +45$^{\circ}$, e:Subimage at 75$^{\circ}$,f: Subimage at -75$^{\circ}$, g: Subimage at -45$^{\circ}$, h:Subimage at -15$^{\circ}$)}
	
	\label{figure4}
\end{figure*}

Using the 2D-DWT an image $I(k_1,k_2)$ can be decomposed as follows\cite{tsiaparas2009discrete}:
\begin{equation}
	\sum_{k_{1}=0}^{N}\sum_{k_{2}=0}^{M}I(k_{1},k_{2})f^{d}_{j}(k_{1}-n_{1},k_{2}-n_{2})=
	\left\{
	\begin{array}{ll}
		A_{j} & if\ d=0 \\
		Dh_{j} & if\ d=1 \\
		Dv_{j} & if\ d=2 \\
		Dd_{j} & if\ d=3 \\
	\end{array} 
	\right.
\end{equation}
Where $N$ and $M$ are the row and column numbers. $A_{j}$ are the approximation coefficients or approximation sub-images and $Dh_{j}$, $Dv_{j}$ and $Dd_{j}$ are the detail coefficients or detail subimages for every level of decomposition j. The function $f^{d}_{j}$ is defined as:
\begin{equation}
	f^{d}_{j}=
	\left\{
	\begin{array}{ll}
		\phi_{j}^*(n_{1})\phi_{j}^*(n_{2}) &  d=0 \\
		\phi_{j}^*(n_{1})\psi_{j}^*(n_{2}) &  d=1 \\
		\psi_{j}^*(n_{1})\phi_{j}^*(n_{2}) &  d=2 \\
		\psi_{j}^*(n_{1})\psi_{j}^*(n_{2}) &  d=3 \\
	\end{array} 
	\right.
\end{equation}

However, an issue with the DWT is that it is shift-variant. This means that if the morphologies in the LUS images were translated, the DWT would generate a different set of DWT coefficients. 

\subsubsection{DTCWT}

In order to address the lack of shift-invariance, DTCWT type approaches can be used as it is nearly shift invariant. Also, DTCWT has good directional selectivity and perfect reconstruction. The 2-D DTCWT uses two separate decomposition trees to calculate the complex transform of an image. One of the trees are used to calculate the real parts of the complex coefficients and the other tree is used to calculate the imaginary parts of the complex coefficients \cite{DTCWT_Kingsbury}.  The DTCWT is implemented by using two separate two-channel filter banks. Approximate shift invariance was achieved by doubling the sampling rate
at each level of the tree by using two trees \cite{DTCWT_Kingsbury}. DTCWT decomposition of an image is done by using a complex scaling function and six wavelet functions \cite{DTCWT_and_GLCM,DTCWT_Kingsbury}. The DTCWT results in a low-passed version of the image at each decomposition level and six high-frequency sub-images at each decomposition level corresponding to the six wavelet functions oriented at angles $\alpha$=($\pm$15$^{\circ}$,$\pm$45$^{\circ}$,$\pm$75$^{\circ}$).

The decomposition of an image $I(k_1,k_2)$ can be performed by using a complex scaling function and six complex wavelet function as follows \cite{DTCWT_Formula}: 

\begin{equation} I(k_1,k_2)=\sum \limits _{l\in Z^{2}} {A_{j_{0},l} \phi _{j_{0},l} (k_1,k_2)}+ \sum \limits _{g\in \alpha } {\sum \limits _{j=1}^{j_{0} } {\sum \limits _{l\in Z^{2}} {D_{j,l}^{g} } } } \psi _{j,l}^{g} (k_1,k_2)\quad 
\end{equation}

Where $j_0$ is the number of decomposition level, $A_{j0,l}$ and $D_{j,l}^{g}$ are scaling coefficients and wavelet coefficients respectively. $\phi_{j_{0},l}(k_1,k_2)$ represents the scaling function and $\psi_{j,l}^g(k_1,k_2)$ represents the six wavelet functions \cite{DTCWT_Formula}.

\subsection{Feature Extraction}

We extracted features from the top and bottom half of the DTCWT sub-images. The rational being, the top half of the images will have features related to the pleural line and the bottom half will have features related to B-lines and A-lines. To extract features that can describe the intensity distribution of the pixels we used statistical features directly on the image, grey-level co-occurrence features and rotation invariant uniform LBP features. To extract features that measure the morphology in the image, we used grey-level run length matrix features. A combination of textural and morphological features was selected to obtain a complementary feature set that should produce a more robust classification model \cite{Stat}. Statistical features were extracted from the pixel intensities of the images \cite{Stat_formulas}.

\subsubsection{Statistical Features}

The below statistical features were chosen as they are commonly used as global features in medical image analysis.

\begin{equation}
	Mean=\frac{\sum_{i=1}^{n} \sum_{j=1}^{m} (I_{x,y})}{m\cdot n}
\end{equation}
\begin{equation}
	SD=\frac{\sqrt{\sum_{i=1}^{n} \sum_{j=1}^{m} (I_{x,y}-Mean)^2}}{m\cdot n}
\end{equation}
\begin{equation}
	Skewness=\frac{\sum_{i=1}^{n} \sum_{j=1}^{m} (I_{x,y}-Mean)^3}{m\cdot n\cdot SD^3}
\end{equation}
\begin{equation}
	Kurtosis=\frac{\sum_{i=1}^{n} \sum_{j=1}^{m} (I_{x,y}-Mean)^4}{m\cdot n\cdot SD^4} -3
\end{equation}
\begin{equation}
	Entropy=-\sum_{i=1}^{n} \sum_{j=1}^{m} I_{x,y}log_{2}(I_{x,y})
\end{equation}

Where $n$ is the number of rows in the image, $m$ is the number of columns in the image, $i$ is the row number, $j$ is the column number and $I$ is the image. 

\subsubsection{GLCM Features}

The grey-level co-occurrence matrix is calculated by determining how often pairs of pixels with specific values and in a specified spatial relationship occur in an image. Then statistical measures are extracted from the GLCM \cite{GLCM}. To calculate the GLCM we first quantized the image to 8 levels, generating an 8x8 GLCM and choose 6 offsets, [0 1; 1 0; 0 2; 2 0; 1 1; 2 2;], to generate 6 different GLCMs. Element $(i,j)$ of the gray level co-occurrence matrix represents the number of occasions a pixel with intensity $i$ is adjacent to a pixel with intensity $j$ in the LUS image. The GLCM stores the co-occurring values representing the distance and angular spatial relationship of pixels in a structured matrix.  We then extracted 5 features from the GLCM \cite{GLCM_formula}. \\

\begin{equation}
	Contrast=\sum_{i=1}^{n} \sum_{j=1}^{m}(i-j)^2 (G_{i,j})
\end{equation}
\begin{equation}
	Correlation=\sum_{i=1}^{n} \sum_{j=1}^{m} \frac{(i-u)(j-u)(G_{i,j})}{\sigma^2}
\end{equation}
\begin{equation}
	Energy=\sum_{i=1}^{n} \sum_{j=1}^{m} {(G_{i,j})^2}
\end{equation}
\begin{equation}
	Homogeniety=\sum_{i=1}^{n} \sum_{j=1}^{m} \frac{(G_{i,j})}{1+(i-j)^2}
\end{equation}
\begin{equation}
	Entropy=-\sum_{i=1}^{n} \sum_{j=1}^{m} (G_{i,j}log_2(G_{i,j})
\end{equation}

Where $i$ is the row number of the GLCM, $j$ is the column number of the GLCM and $G$ is the GLCM.
Where $u$ is the mean pixel intensity in the quantized image and $\sigma^2$ is the variance of the pixel intensity in the quantized image.
\subsubsection{GLRLM Features}

The grey-level run length matrix is used to store run lengths based on grey level value and length of the run. A grey level run is a set of pixels having the same grey level value, pixels are consecutive and collinearly distributed in some given direction \cite{GLRLM}. We calculated the 4 GLRLM for each of the 2 ROIs in the image using 4 directions 0$^{\circ}$, 45$^{\circ}$, 90$^{\circ}$ and 135$^{\circ}$. Then extracted 11 features from the GLRLM. The equations are shown in Table \ref{table3} \cite{GLRLM_formula, GLRLM_formula_journal}.

\begin{table*}[!h]
	\centering
	\begin{threeparttable}  
		\caption{GLRLM Equations}
		\label{table3}
		\centering
		\begin{tabular}{|l|l|}
			\hline
			\(\displaystyle  	LRE=\frac{\sum_{i=1}^{n} \sum_{j=1}^{m}j^2 (R_{i,j})}{\sum_{i=1}^{n} \sum_{j=1}^{m} R_{i,j}} 	\qquad{(14)}\)   &  \(\displaystyle SRE=\frac{\sum_{i=1}^{n} \sum_{j=1}^{m} \frac{(R_{i,j})}{j^2}}{\sum_{i=1}^{n} \sum_{j=1}^{m} R_{i,j}} \quad{(15)}\)\\ \hline
			\(\displaystyle GLN=\frac{\sum_{i=1}^{n} (\sum_{j=1}^{m} (R_{i,j})^2)}{\sum_{i=1}^{n} \sum_{j=1}^{m} R_{i,j}} \quad{(16)}\) &  \(\displaystyle RLN=\frac{\sum_{j=1}^{m} (\sum_{i=1}^{n} (R_{i,j})^2)}{\sum_{i=1}^{n} \sum_{j=1}^{m} R_{i,j}}  \quad{(17)}\)\\ \hline
			\(\displaystyle RP=\frac{\sum_{i=1}^{n} (\sum_{j=1}^{m} R_{i,j})}{N} \qquad{(18)}\) &  \(\displaystyle LGRE=\frac{\sum_{i=1}^{n} \sum_{j=1}^{m} \frac{(R_{i,j})}{i^2}}{\sum_{i=1}^{n} \sum_{j=1}^{m} R_{i,j}} \quad{(19)}\)\\ \hline
			\(\displaystyle HGRE=\frac{\sum_{i=1}^{n} \sum_{j=1}^{m}i^2 (R_{i,j})}{\sum_{i=1}^{n} \sum_{j=1}^{m} R_{i,j}} \qquad{(20)}\) &  \(\displaystyle SRLGE=\frac{\sum_{i=1}^{n} \sum_{j=1}^{m} \frac{(R_{i,j})}{i^2\cdot j^2}}{\sum_{i=1}^{n} \sum_{j=1}^{m} R_{i,j}} \quad{(21)}\)\\ \hline
			\(\displaystyle SRHGE=\frac{\sum_{i=1}^{n} \sum_{j=1}^{m} \frac{i^2(R_{i,j})}{j^2}}{\sum_{i=1}^{n} \sum_{j=1}^{m} R_{i,j}} \qquad{(22)}\) &  \(\displaystyle LRLGE=\frac{\sum_{i=1}^{n} \sum_{j=1}^{m} \frac{j^2(R_{i,j})}{i^2}}{\sum_{i=1}^{n} \sum_{j=1}^{m} R_{i,j}} \quad{(23)}\)\\ \hline
		\end{tabular}
		\begin{tabular}{|ll|}	
				\(\displaystyle \hspace{27 mm} LRHGE=\frac{\sum_{i=1}^{n} \sum_{j=1}^{m} i^2\cdot j^2 (R_{i,j})}{\sum_{i=1}^{n} \sum_{j=1}^{m} R_{i,j}} \quad{(24)}\) & \(\displaystyle \hspace{27 mm} \)\\\hline 
			
		\end{tabular}
		\begin{tablenotes}
			\small{
				\item *In the above GLRLM feature equations, $i$ is the grey level, $j$ is the run length and $R$ is the GLRLM. }
		\end{tablenotes}
	\end{threeparttable}  
	\vspace{0.1in}	
\end{table*}

\subsubsection{LBP Features}

The LBP histogram is used to store LBP pixel labels for the image. The pixel labels are calculated by thresholding the 3x3-neighborhood of each pixel with the center value and considering the result as a binary number \cite{LBP}. We used the rotation-invariant LBP which results in 10 bins. First, only uniform LBPs, patterns with a maximum of two circular 0-1 or 1-0 transitions are considered unique patterns, all nonuniform patterns are stored in one bin. Then, in the rotation-invariant version LBP patterns that result in the same value when rotated are stored in the same bin. The values of the bins were used as the feature set. 

\subsubsection{Clinical Features}

As with our previous initial work and as explained earlier, we included 3 clinical features (GA, CGATS, and DOL) in our models as they are essential in segregating CLD and RDS. For the diseases CLD and RDS, the LUS images morphologies are not enough to separate the conditions as affirmed by our clinical collaborators. Without these features a meaningful clinical diagnosis of the conditions is not possible or extremely difficult only using LUS images. While these are important features in a classification sense, it is critical to note that the clinical features themselves cannot separate the conditions very well. For example, Normal and PTX are unrelated to the clinical features and can occur in neonates regardless of gestation age, days of life and age at the time of scan. The clinical features can only separate the conditions in a meaningful way and play a supportive role when LUS information is made available.

\subsection{Pattern Classification}

In this work, we used a simple Linear Discriminant Analysis (LDA) based classifier to perform the classification of the 6 LUS conditions using the features extracted from the LUS images. While performing classification on the balanced dataset, the models were trained based on equal prior probabilities between the groups and for the whole dataset we used prior probabilities based on group size. Simple linear classifier was chosen over the complex non-linear classifiers to place the emphasis more on extracting meaningful features relevant to clinical markers and to keep the outcomes conservative and realistic. Likewise, we used an univariate feature selection procedure using chi-square tests \cite{perez2014improving} due its speed as a low computational complexity feature selection method was required for doing feature selection inside the loop of cross-validation. In chi-square feature selection process, a individual chi-square test is performed for each feature and the class labels. A small p-value indicates that the corresponding feature is dependent on the class, and is ranked accordingly as an important feature.

In terms of cross validation, we tested the performance using two forms of cross-validation. We used  leave-one-out cross-validation (LOO CV), which is a form of cross-validation where only one image is used as the testing set and the model is trained on the rest of the images. This is repeated so that each of the images is used as a testing set. The average accuracy of the all these runs then results in per-image classification performance. We also used leave-one-subject-out cross-validation (LOSO CV). In this form of cross-validation, all of the images belonging to a subject are used as the testing set and the rest of the images are used as the training set. This is repeated until each subject has served in the testing set. The average accuracy of the all these runs then results in per-subject classification performance. The motivation behind the LOSO cross-validation is to avoid biasing the model due to the similarity between images of the same patient. We also only selected 5 images from each video to avoid the large similarity between frames in the same video for our LOO results. The images were selected at equal intervals throughout the video. If the lung was not clearly imaged in the frame another frame was selected as was the case at the beginning or end of a few videos. 

\begin{table*}[h!]
	\centering
	\caption{Classification confusion matrix for results with LOO CV on BALANCED dataset using top 15 DTCWT features and 3 clinical features. The overall per-image classification accuracy achieved is {\bf 92.78\%} }
	\begin{tabular}{|c|c|cccccc|}
		\hline
		&        & \multicolumn{6}{c|}{Predicted Class}                                                                                                                            \\ \hline
		&        & \multicolumn{1}{c|}{Normal}  & \multicolumn{1}{c|}{CLD}   & \multicolumn{1}{c|}{CON}     & \multicolumn{1}{c|}{PTX}     & \multicolumn{1}{c|}{RDS}    & TTN     \\ \hline
		\multirow{6}{*}{\begin{tabular}[c]{@{}c@{}}True\\ Class\end{tabular}} & Normal & \multicolumn{1}{c|}{\bf 98.33\%} & \multicolumn{1}{c|}{0\%}   & \multicolumn{1}{c|}{0\%}     & \multicolumn{1}{c|}{0.83\%}  & \multicolumn{1}{c|}{0\%}    & 0.83\%  \\ \cline{2-8} 
		& CLD    & \multicolumn{1}{c|}{0\%}     & \multicolumn{1}{c|}{\bf 100\%} & \multicolumn{1}{c|}{0\%}     & \multicolumn{1}{c|}{0\%}     & \multicolumn{1}{c|}{0\%}    & 0\%     \\ \cline{2-8} 
		& CON    & \multicolumn{1}{c|}{0\%}     & \multicolumn{1}{c|}{0\%}   & \multicolumn{1}{c|}{\bf 95.83\%} & \multicolumn{1}{c|}{0\%}     & \multicolumn{1}{c|}{4.17\%} & 0\%     \\ \cline{2-8} 
		& PTX    & \multicolumn{1}{c|}{9.17\%}  & \multicolumn{1}{c|}{0\%}   & \multicolumn{1}{c|}{0.83\%}  & \multicolumn{1}{c|}{\bf 76.67\%} & \multicolumn{1}{c|}{0\%}    & 13.33\% \\ \cline{2-8} 
		& RDS    & \multicolumn{1}{c|}{0\%}     & \multicolumn{1}{c|}{0\%}   & \multicolumn{1}{c|}{4.17\%}  & \multicolumn{1}{c|}{0\%}     & \multicolumn{1}{c|}{\bf 92.5\%} & 3.33\%  \\ \cline{2-8} 
		& TTN    & \multicolumn{1}{c|}{3.33\%}  & \multicolumn{1}{c|}{0\%}   & \multicolumn{1}{c|}{0\%}     & \multicolumn{1}{c|}{3.33\%}  & \multicolumn{1}{c|}{0\%}    & {\bf 93.33\%} \\ \hline
	\end{tabular}
	\vspace{0.2cm}
	\label{T1}
\end{table*}

\begin{table*}[h!]
	\centering
	\caption{Classification confusion matrix for results with LOO CV on WHOLE dataset using top 15 DTCWT features and 3 clinical features. The overall per-image classification accuracy achieved is {\bf 74.39\%}}
	\begin{tabular}{|c|c|cccccc|}
		\hline
		&        & \multicolumn{6}{c|}{Predicted Class}                                                                                                                             \\ \hline
		&        & \multicolumn{1}{c|}{Normal}  & \multicolumn{1}{c|}{CLD}   & \multicolumn{1}{c|}{CON}     & \multicolumn{1}{c|}{PTX}     & \multicolumn{1}{c|}{RDS}     & TTN     \\ \hline
		\multirow{6}{*}{\begin{tabular}[c]{@{}c@{}}True\\ Class\end{tabular}} & Normal & \multicolumn{1}{c|}{\bf 17.84\%} & \multicolumn{1}{c|}{0\%}   & \multicolumn{1}{c|}{0\%}     & \multicolumn{1}{c|}{7.03\%}  & \multicolumn{1}{c|}{0\%}     & 75.14\% \\ \cline{2-8} 
		& CLD    & \multicolumn{1}{c|}{0\%}     & \multicolumn{1}{c|}{\bf 100\%} & \multicolumn{1}{c|}{0\%}     & \multicolumn{1}{c|}{0\%}     & \multicolumn{1}{c|}{0\%}     & 0\%     \\ \cline{2-8} 
		& CON    & \multicolumn{1}{c|}{0\%}     & \multicolumn{1}{c|}{0\%}   & \multicolumn{1}{c|}{\bf 99.02\%} & \multicolumn{1}{c|}{0\%}     & \multicolumn{1}{c|}{0.98\%}  & 0\%     \\ \cline{2-8} 
		& PTX    & \multicolumn{1}{c|}{8.57\%}  & \multicolumn{1}{c|}{0\%}   & \multicolumn{1}{c|}{12.38\%} & \multicolumn{1}{c|}{\bf 17.14\%} & \multicolumn{1}{c|}{25.71\%} & 36.19\% \\ \cline{2-8} 
		& RDS    & \multicolumn{1}{c|}{2.86\%}  & \multicolumn{1}{c|}{0\%}   & \multicolumn{1}{c|}{8.98\%}  & \multicolumn{1}{c|}{5.31\%}  & \multicolumn{1}{c|}{\bf 57.55\%} & 25.31\% \\ \cline{2-8} 
		& TTN    & \multicolumn{1}{c|}{4.53\%}  & \multicolumn{1}{c|}{0\%}   & \multicolumn{1}{c|}{0\%}     & \multicolumn{1}{c|}{2.26\%}  & \multicolumn{1}{c|}{2.83\%}  & \bf 90.38\% \\ \hline
	\end{tabular}
	\vspace{0.2cm}
	\label{T2}
\end{table*}

\begin{table*}[h!]
	\centering
	\caption{Classification confusion matrix for results with LOSO CV on BALANCED dataset using top 15 DTCWT features and 3 clinical features. The overall per-subject classification accuracy achieved is {\bf 75\%}}
	\begin{tabular}{|c|c|cccccc|}
		\hline
		&        & \multicolumn{6}{c|}{Predicted Class}                                                                                                                             \\ \hline
		&        & \multicolumn{1}{c|}{Normal}  & \multicolumn{1}{c|}{CLD}   & \multicolumn{1}{c|}{CON}     & \multicolumn{1}{c|}{PTX}     & \multicolumn{1}{c|}{RDS}     & TTN     \\ \hline
		\multirow{6}{*}{\begin{tabular}[c]{@{}c@{}}True\\ Class\end{tabular}} & Normal & \multicolumn{1}{c|}{\bf 93.33\%} & \multicolumn{1}{c|}{0\%}   & \multicolumn{1}{c|}{0\%}     & \multicolumn{1}{c|}{1.67\%}  & \multicolumn{1}{c|}{0\%}     & 5\%     \\ \cline{2-8} 
		& CLD    & \multicolumn{1}{c|}{0\%}     & \multicolumn{1}{c|}{\bf 100\%} & \multicolumn{1}{c|}{0\%}     & \multicolumn{1}{c|}{0\%}     & \multicolumn{1}{c|}{0\%}     & 0\%     \\ \cline{2-8} 
		& CON    & \multicolumn{1}{c|}{0\%}     & \multicolumn{1}{c|}{0\%}   & \multicolumn{1}{c|}{\bf 78.33\%} & \multicolumn{1}{c|}{0\%}     & \multicolumn{1}{c|}{21.67\%} & 0\%     \\ \cline{2-8} 
		& PTX    & \multicolumn{1}{c|}{11.67\%} & \multicolumn{1}{c|}{0\%}   & \multicolumn{1}{c|}{2.5\%}   & \multicolumn{1}{c|}{\bf 40.83\%} & \multicolumn{1}{c|}{5.83\%}  & 39.17\% \\ \cline{2-8} 
		& RDS    & \multicolumn{1}{c|}{0\%}     & \multicolumn{1}{c|}{0\%}   & \multicolumn{1}{c|}{28.33\%} & \multicolumn{1}{c|}{0\%}     & \multicolumn{1}{c|}{\bf 52.5\%}  & 19.17\% \\ \cline{2-8} 
		& TTN    & \multicolumn{1}{c|}{9.17\%}  & \multicolumn{1}{c|}{0\%}   & \multicolumn{1}{c|}{0\%}     & \multicolumn{1}{c|}{5.83\%}  & \multicolumn{1}{c|}{0\%}     & \bf 85\%    \\ \hline
	\end{tabular}
	\vspace{0.2cm}
	\label{T3}
\end{table*}

\begin{table*}[h!]
	\centering
	\caption{Classification confusion matrix for results with LOSO CV on WHOLE dataset using top 15 DTCWT features and 3 clinical features. The overall per-subject  classification accuracy achieved is {\bf 63.48\%}}
	\begin{tabular}{|c|c|cccccc|}
		\hline
		&        & \multicolumn{6}{c|}{Predicted Class}                                                                                                                             \\ \hline
		&        & \multicolumn{1}{c|}{Normal}  & \multicolumn{1}{c|}{CLD}   & \multicolumn{1}{c|}{CON}     & \multicolumn{1}{c|}{PTX}     & \multicolumn{1}{c|}{RDS}     & TTN     \\ \hline
		\multirow{6}{*}{\begin{tabular}[c]{@{}c@{}}True\\ Class\end{tabular}} & Normal & \multicolumn{1}{c|}{\bf 4.86\%}   & \multicolumn{1}{c|}{0\%}   & \multicolumn{1}{c|}{0\%}     & \multicolumn{1}{c|}{11.89\%} & \multicolumn{1}{c|}{0\%}     & 83.24\% \\ \cline{2-8} 
		& CLD    & \multicolumn{1}{c|}{0\%}     & \multicolumn{1}{c|}{\bf 100\%} & \multicolumn{1}{c|}{0\%}     & \multicolumn{1}{c|}{0\%}     & \multicolumn{1}{c|}{0\%}     & 0\%     \\ \cline{2-8} 
		& CON    & \multicolumn{1}{c|}{0\%}     & \multicolumn{1}{c|}{0\%}   & \multicolumn{1}{c|}{\bf 96.39\%} & \multicolumn{1}{c|}{0\%}     & \multicolumn{1}{c|}{3.61\%} & 0\%  \\ \cline{2-8} 
		& PTX    & \multicolumn{1}{c|}{23.81\%} & \multicolumn{1}{c|}{0\%}   & \multicolumn{1}{c|}{15.24\%} & \multicolumn{1}{c|}{\bf 3.81\%}  & \multicolumn{1}{c|}{22.86\%} & 34.29\%    \\ \cline{2-8} 
		& RDS    & \multicolumn{1}{c|}{5.71\%}  & \multicolumn{1}{c|}{0\%}   & \multicolumn{1}{c|}{17.96\%} & \multicolumn{1}{c|}{6.12\%}  & \multicolumn{1}{c|}{\bf 39.59\%} & 30.61\% \\ \cline{2-8} 
		& TTN    & \multicolumn{1}{c|}{13.96\%} & \multicolumn{1}{c|}{0\%}   & \multicolumn{1}{c|}{0\%}     & \multicolumn{1}{c|}{5.09\%}  & \multicolumn{1}{c|}{5.47\%}  & \bf 75.47\% \\ \hline
	\end{tabular}
	\label{T4}
\end{table*}

\begin{table*}[h!]
	\centering
	\caption{Maximum accuracy obtained using DTCWT features and 3 clinical features with LOSO CV on the balanced and the whole dataset (Number of features given in brackets)}
	\begin{tabular}{|c|c|c|c|c|}
		\hline
		& \begin{tabular}[c]{@{}c@{}} Balanced Dataset\end{tabular} & \begin{tabular}[c]{@{}c@{}}Whole Dataset\end{tabular} \\ \hline
		LOSO CV Accuracy & {{\bf81.53\%}  (46)}                                               & {{\bf 64.97\%}  (16) }\\ \hline
	\end{tabular}
	\label{LOSO_feat_dim}
\end{table*}

\section{RESULTS AND DISCUSSION}

We performed the following four 6-group classification experiments: {\bf (i)} LDA with LOO CV on balanced dataset, {\bf (ii)} LDA with LOO CV on the whole dataset, {\bf (iii)} LDA with LOSO CV on balanced dataset  and {\bf (iv)} LDA and LOSO CV on the whole dataset. In performing the above experiments we also restricted the feature dimension to the top 15 DTCWT features selected by the feature selection method and 3 clinical features. Considering the approximate training group sample sizes, 18 features will form ~10-15\% of the training group sizes which is expected to produce conservative/ realistic performance without lose of generalization or over-fitting. We also tested the distribution of top selected features by the feature selection method and found that in general GLCM and GLRM features were selected in larger proportions in comparison to statistical and LBP features.  

The obtained results are presented in the four confusion matrices in Tables \ref{T1}-\ref{T4}. Table \ref{T1} presents the results of the 6-group classification with LOO CV on the balanced dataset. With an overall per-image classification accuracy of {\bf 92.78\%}, most of the groups are classified well expect PTX which has a slight overlap between Normal and TTN. Table \ref{T2} presents the results of the 6-group classification with LOO CV on the whole dataset. With an overall per-image classification accuracy of {\bf 74.39\%}, the performance drops considerably in comparison to the balanced dataset. As evident from the Table \ref{T2}, looking at the TTN column most of the issue seems to be the overlap with TTN. This is expected as the whole dataset is skewed due to almost 1/3rd of TTN cases and also as confirmed by our clinical collaborator, TTN can have varying and overlapping presentation with Normal, PTX and RDS. Specifically, for the whole dataset experiment, Normal was misclassified as TNN 75.14\% of the time, PTX was misclassified as TTN 36.19\% of the time, and RDS was misclassiifed as TTN 25.31\% of the time. These 3 conditions look very similar, they all commonly have A-lines, however there are differences between these diseases such as the the amount of fluid present in the lungs  which can be picked up in the balanced dataset. These might have contributed to the reduced performance of the per-image classification using the whole dataset.

\begin{figure}
	\vspace{-0.8cm}
	\centering
	\includegraphics[trim=3.75cm 8.5cm 2cm 7cm,clip=true,width=10cm]{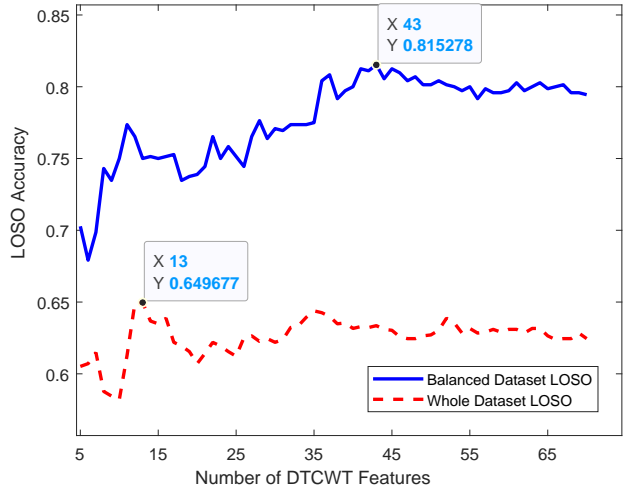}
	\caption{LOSO Accuracies for Balanced and Whole Dataset with increased Number of DTCWT Features}
	\label{f1}
\end{figure}

Moving on to the LOSO results in Table \ref{T3} and \ref{T4}, we see in general a reduction as expected in the per-subject classification accuracies in comparison with the per-image classification accuracies in Table \ref{T1}  and \ref{T2}. This could be explained due to two reasons. First, we have removed any subject bias that may have influenced the results using LOO CV with images. Secondly, LOSO CV reduces the amount of condition specific data the model is trained on for the patient in the testing set, which affects the accuracy of the model, which could lead to a decrease in accuracy. In addition, for the whole dataset case, the unbalanced and 1/3rd presence of the TTN might have amplified the difficulty level of classifying the groups. Lastly the reduced feature dimension might not have enough discriminative ability to separate the groups well.

To test whether increasing the features dimension beyond top 15 DTCWT features and 3 clinical features could potentially improve the classification abilities, we repeated the LOSO classification experiments with both the balanced and the whole dataset. In Fig. \ref{f1} we have illustrated the the change in the LOSO overall accuracy with the increase in the number of DTCWT features. The datatips in the figure only show the number of DTCWT features (X axis) but the LOSO accuracy on the Y axis was computed with the combination of DTCWT features and the 3 clinical features. As evident from the figure a maximum overall accuracy of {\bf 81.53\%} is achieved with the top 43 DTCWT features and 3 clinical features for the balanced dataset, however the improvement for the whole dataset was minimal. The improved overall classification accuracies with corresponding DTCWT feature dimensions are provided in Table \ref{LOSO_feat_dim}. This alludes that there may be scope for improvement with optimization in the feature dimension (or) using advanced machine learning methods which is out of the scope of this paper.    

As reported in the introduction, most of the existing works around LUS only try to detect a single pathological disease. This makes it difficult to compare our DTCWT feature extraction algorithm with existing works as our work is more comprehensive performing a multi-group classification with a larger dataset. In another previous work from our group \cite{c10}, a smaller but comparable dataset containing images from 5 conditions was used. That work used an object detection model to detect the 7 LUS morphologies, but did not perform direct pathological condition classification. The object detection model does produce meaningful outputs, which are comparable to the meaningful image decomposition features captured in this work. In addition, the proposed work performed pathological condition classification and included 3 clinical features which enabled us to separate certain lung conditions that are inseparable from images alone as verified by our clinical collaborators. To compare our previous initial work using recurrence features \cite{RQA_Paper} with DTCWT features, we recomputed the results using recurrence features with updated datasets used in this work. We were able to achieve a per-image classification accuracy of {\bf 85.42\%} on the balanced dataset with LOO CV and {\bf 72.00\%} with LOO CV on the whole dataset using the recurrence features. These results were still lower than the proposed DTCWT work.

The results and discussion provided above demonstrates the potential of using a DTCWT based approach to build an automated machine learning systems that could eventually lead to decision support systems to assist the clinical community. While there is scope for improvement especially in TTN related misclassifications albeit high LOO CV accuracies, the promising potential of designing a system that could serve as a first level screening tool to identify lung conditions from LUS images is encouraging. Especially in remote communities and developing countries with lack of specialist clinicians such system could help with timely diagnosis and treatment of neonates suffering respiratory diseases, thereby saving lives. 

\section{CONCLUSIONS}

To facilitate the widespread use of LUS an automated classification tool could be used to assist medical professionals and physicians in hospitals where there is a lack of highly trained medical professionals and clinicians. In this work, we have demonstrated that simple DTCWT features extracted from LUS images can be used in a system to classify the six common LUS pathologies in neonates. With a LDA based classifier, the proposed classification model achieved a LOO CV per-image classification accuracy of {\bf 92.78\%} on the balanced dataset and a LOO CV per-image classification accuracy of {\bf 74.39\%} on the whole dataset. Likewise the proposed approach achieved a maximum LOSO CV per-subject classification accuracy of {\bf 81.53\%} on the balanced dataset and a maximum LOSO CV per-subject classification accuracy of {\bf 64.97\%} on the whole dataset. The proposed DTCWT features along with the 3 clinical features performed fairly well at separating the 6 lung pathologies. 

Future work involves extracting dynamic features to detect lung sliding from post-processed M-mode images to help separate Normal, PTX and TTN. Our team is also working on advanced machine learning, deep learning approaches, and feature fusion techniques which will augment these results helping to achieve better robustness and reliability. 

\section*{ACKNOWLEDGMENT}

We would like to acknowledge the Natural Sciences and Engineering Research Council of Canada (NSERC)-via Alliance Grant (ALLRP 546302-19) for providing the resources to make this work possible. Additionally, we would like to acknowledge Jenna Ibrahim (Mount Sinai Hospital) for her work in helping to create the database and facilitating the meetings with our clinical collaborators.

\bibliographystyle{unsrt}
\bibliography{References_DTCWT}
	
\end{document}